# Experimental investigation of water emulsion fuel stability

**Gurjap Singh[1]**
University of Iowa
Iowa City, IA, United States

**Elio Lopes**
Santa Catarina State University
Joinville, Brazil

**Nicholas Hentges**
University of Iowa
Iowa City, IA, United States

**Albert Ratner**
University of Iowa
Iowa City, IA, United States

**ABSTRACT**

The combustion of liquid fuels emulsified with water have long generated interest in the internal combustion engine research community. Typically, these fuels consist of small quantities of water emulsified with ultrasonification or other mechanical methods into a pure or multicomponent hydrocarbon fuel. These emulsion fuels promise significant advantages over base liquid fuels, such as better fuel economy, colder combustion temperatures, less NOx emissions, and so on. However, a significant practical disadvantage of these fuels is that they are prone to phase separation after they have been prepared. Till date, an objective but economical method of identifying the various degrees of phase separation has not been identified. Present research presents such a method and shows its utilization in analyzing the stability of water and hydrocarbon fuel emulsions over time without the addition of chemical stabilizers. It is expected that present research will pave the way in establishing this method to study the stability of other specialized multicomponent fluids.

Keywords: emulsion stability; stability evaluation; experimental stability evaluation; water-based emulsion; biodiesel-water emulsion;

**NOMENCLATURE**

Place nomenclature section, if needed, here. Nomenclature should be given in a column, like this:

| | |
|---|---|
| BD | biodiesel |
| NOx | oxides of nitrogen |
| PD | petrodiesel |
| SBD | soy biodiesel |

## 1. INTRODUCTION

The emulsification of biodiesel (BD) and petrodiesel (PD) with water and other materials has long been researched [1], [2], [3], [4] for combustion related purposes. Previous research has explored single droplet combustion characteristics for BD and PD [5] and their water emulsions [6]. Many benefits have been associated with emulsifying liquid fuels, such as lesser NOx emissions, leaner combustion, and smaller soot particulate emissions [7]. As PD resources are depleted, the characterization of BD combustion becomes more and more important, therefore the methods to improve these properties need more attention.

A biodiesel-water emulsion is fundamentally a water-in-oil or oil-in-water emulsion. Despite thermodynamic stability of such emulsions, their phase instability cannot be ameliorated without the use of stabilizing agents such as emulsifiers or surfactants [4]. On the other hand, such additives can change important fuel properties like viscosity [8], and neutrally stable fuels are preferable.

Before emulsion-based fuels can be commercially important, it is important to analyze their stability over time and resistance to phase separation. A simple, easy-to-use method is the settling bed method or the bottled test method [9]. It involves filling equal amounts of a given emulsion in a glass bottle or culture tube, and manually noting at given time intervals the amount of separation that has occurred. This method does not require any specialized equipment but is subjective and inconvenient.

Another common method is centrifugal separation [10]. The sample is loaded into culture tubes and spun at a given RPM for a given amount of time (at 3,000 RPM for 5 minutes in the study undertaken by Lin et al. [10]). Whatever sample is the least separated after undergoing the process is the most stable. This method is less time-intensive but requires specialized equipment and is also subjective.

Since both soy biodiesel (SBD) and water are optically transparent, a phase-contrast microscope [11] is also an appropriate tool for analyzing the size distribution of the dispersed phase in an emulsion. However, there is a limitation on the size of the dispersed phase size that can be successfully observed [12].

Present research details the stability analysis of five different SBD-water emulsions using a non-contact, non-invasive method that uses low-cost and off-the-shelf components. The technique yields real-time results that are intuitive and easy to interpret.

## 2. MATERIALS AND METHODS

Emulsions made from soy-based biodiesel (Western Dubuque Biodiesel, Farley, Iowa) and distilled water (CVS Pharmacy, Iowa City, Iowa) were prepared in different ratios by ultrasonication for 30 minutes. A Biologics 3000MP with a 3/16" probe was used for ultrasonicating the sample, which was placed in a 50 ml glass Erlenmeyer flask. The ultrasonicator generated a pulse of 4 seconds, with 4 seconds between two consecutive pulses, to keep heat generation in the sample at a minimum. The prepared sample was immediately transferred to the stability analyzer. **TABLE 1** lists the compositions of all emulsions tested in present work. All samples were prepared and tested at room temperature and pressure for at least four and a half days. No surfactant or stabilizer was used.

**TABLE 1:** COMPOSITION OF FUELS EMULSIONS TESTED

| Name | % w/w Water | % w/w SBD |
|---|---|---|
| 25W | 25 | 75 |
| 40W | 40 | 60 |
| 50W | 50 | 50 |
| 60W | 60 | 40 |

---

[1] Contact author: gurjap-singh@uiowa.edu

| 75W | 75 | 25 |

The stability analyzer used in present research has been detailed previously in the work of Singh et al. [13] [14], where it was used to quantify the stability of nanomaterials suspended in liquid fuels. It is seen that when SBD and water (clear liquids) are sonicated, the resulting emulsion is a milky, turbid liquid. Since it is inherently unstable, both constituent liquids separate out into clear phases over time. Therefore, the opacity of an emulsion changes over time as phase separation occurs, which is the property that is measured by the stability analyzer.

**FIGURE 2** shows a single experimental block of the stability analyzer. A light source (bright white LED, Adafruit) is shined through the sample, that is kept in a standard culture tube or test tube (Fisher Scientific). A phototransistor (Adafruit) directly opposite to the LED is used to measure the amount of light that passes through the sample. The signal generated from the phototransistor is proportional to the amount of light incident on it, and therefore it is proportional to the amount of suspended fraction in the sample. When the sample is completely opaque, the signal generated is 0. When the sample is completely transparent, the signal generated is 1. When dealing with real-life suspensions in the study, 0 corresponds to a fresh, fully suspended emulsion. As phase separation occurs, the signal climbs up to 1.

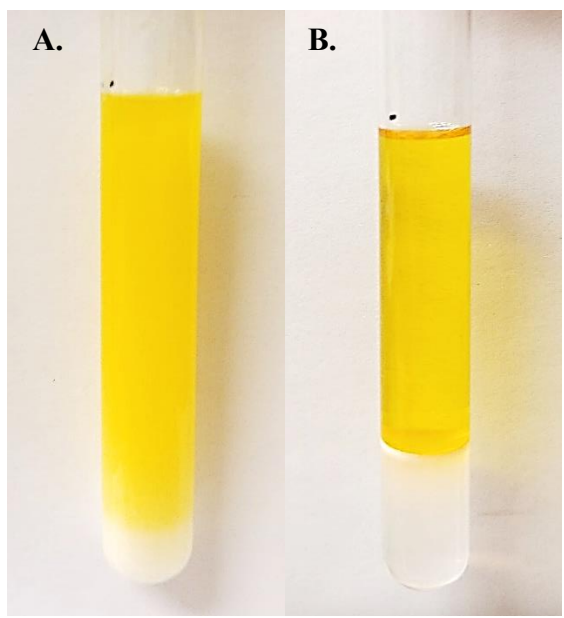

**FIGURE 1:** 25W FUEL EMULSION A. WHEN FRESHLY PREPARED, AND B. AFTER PHASE SEPARATION

Five LED-phototransistor pairs are used at five different levels of the test tube. The stability analyzer can simultaneously analyze up to 30 samples. The signals are collected by a data acquisition card (Arduino Atmega 2560), and the resulting data is logged by a data logger (Raspberry Pi 3 Model B). **FIGURE 3** shows all the major components.

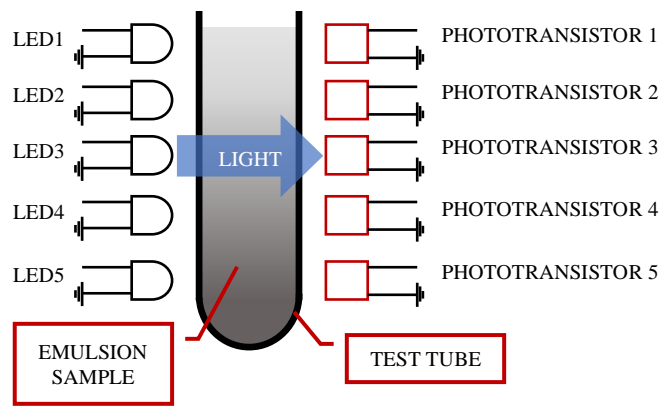

**FIGURE 2:** ONE EXPERIMENTAL BLOCK OF THE SUSPENSION STABILITY ANALYZER SHOWING VARIOUS COMPONENTS

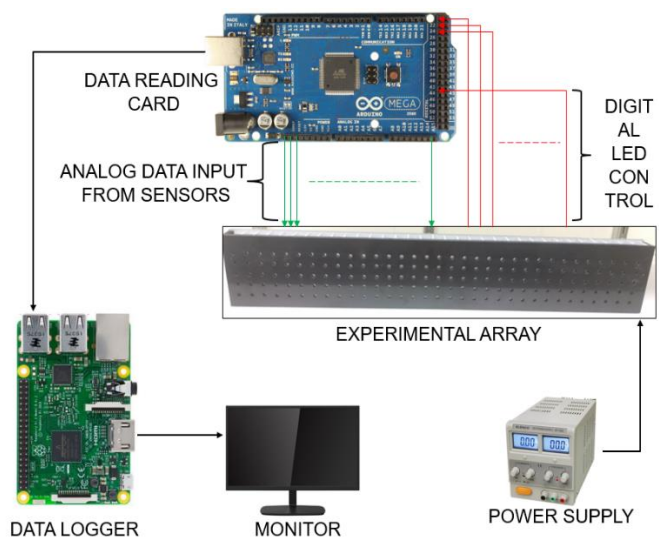

**FIGURE 3:** MAIN COMPONENTS OF THE STABILITY ANALYZER [13]

## 3. RESULTS AND DISCUSSION
### 3.1 General behavior

In previous work that analyzed nanomaterial suspension stability [14], an initial "settling delay" period was observed where the signal stayed at "0". No such behavior was observed for any of the emulsions, which were observed to settle out immediately after they had been prepared. For each sensor, almost 1300 readings were taken, or 6500 total readings for a given sample. All settling trends are provided in **FIGURE 4**, where moving averages (average 20, move 10) method has been used to reduce the data for representation purposes. This reduced data has been used in Section 3.2 for calculating settling trends as well, as will be described later.

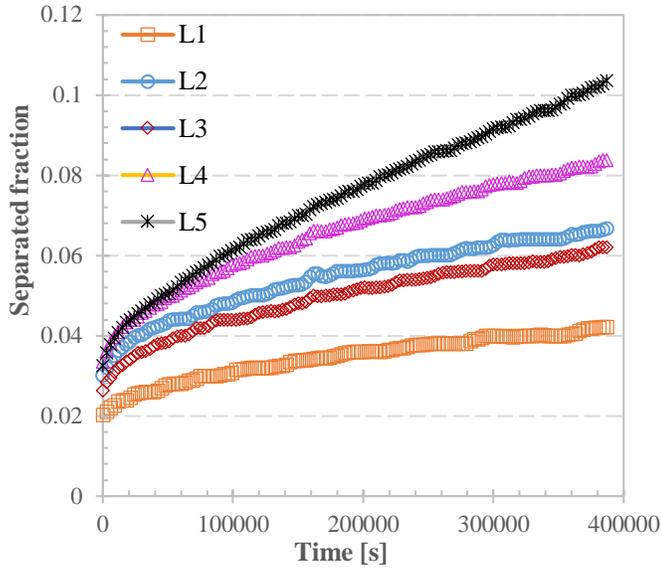

A.

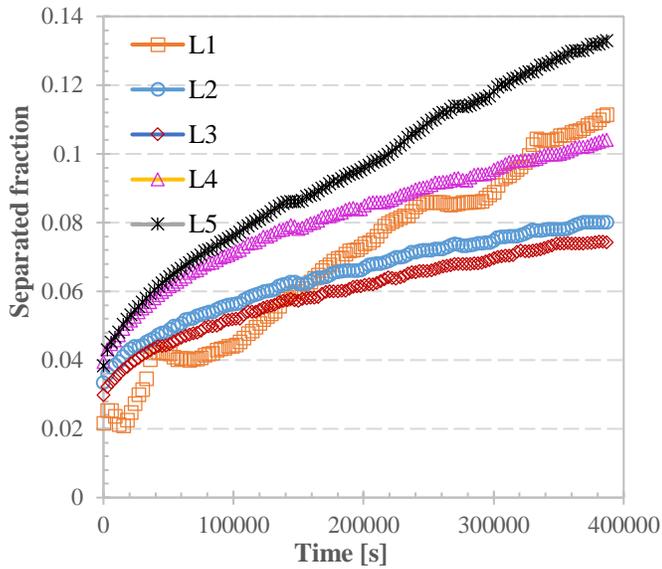

B.

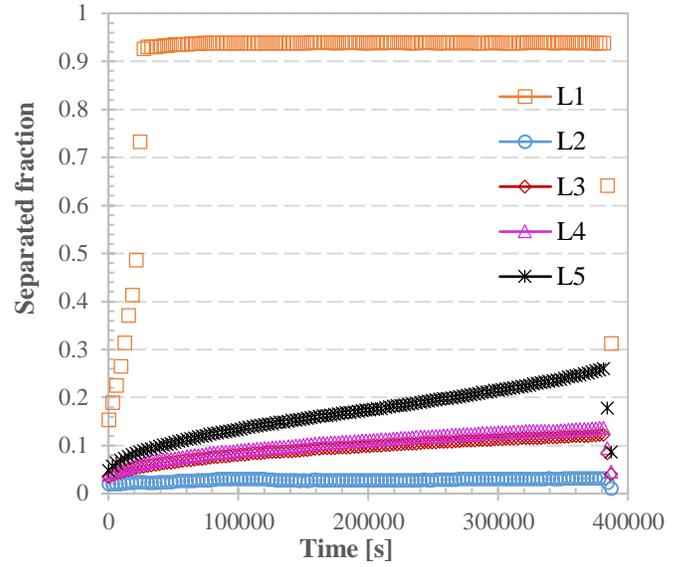

C.

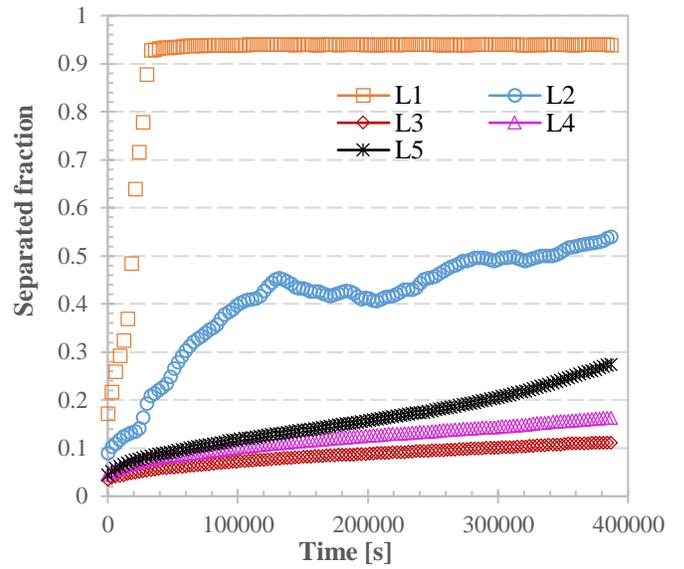

D.

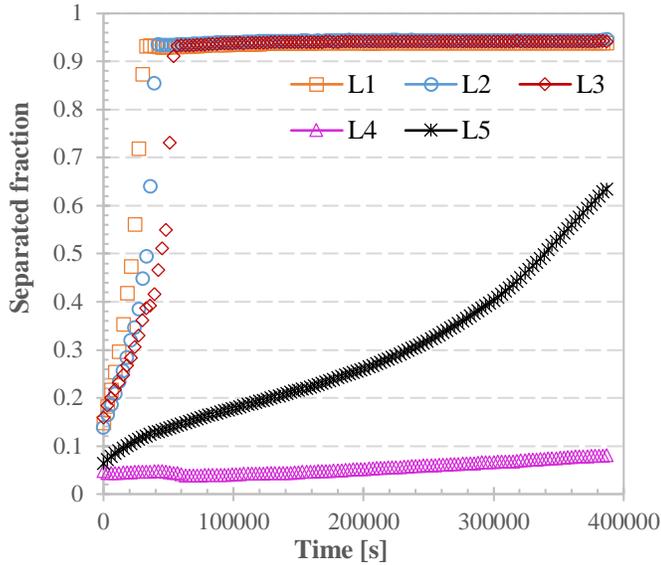

E.
**FIGURE 4:** SEPARATION CHARACTERISTICS OF A. 25W, B. 40W, C. 50W, D. 60W, E. 75W EMULSIONS

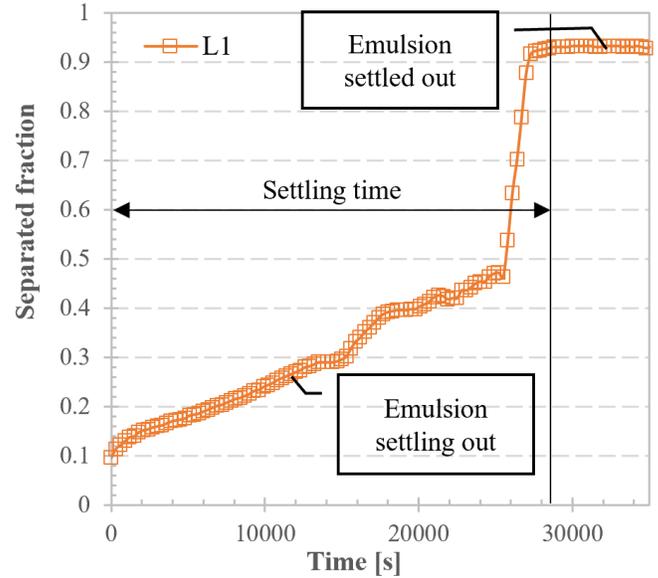

**FIGURE 5:** SEPARATION CHARACTERISTICS OF 50W EMULSION SHOWING SETTLING TIME

### 3.1 Settling time

Settling time is defined as the time after which the emulsion is stable for two hours or more. It was found that 25W and 40W emulsions stayed very stable over the testing period and signal decay was minimal. Meaningful settling time could be calculated only for 50W, 60W and 75W emulsions. Whichever signal decayed the fastest was the one that would indicate actual emulsion stability, and it was found in all cases that L1 signal decayed the fastest.

**TABLE 2:** SETTLING TIMES FOR DIFFERENT EMULSIONS

| Emulsion | Settling time [s] |
|---|---|
| 50W | 29100 |
| 60W | 34500 |
| 75W | 34800 |

Overall, emulsions get more unstable as more water is added. An explanation for that is the higher surface tension of water compared to SBD, which causes a higher surface energy in the dispersed phase and more stability.

### 3.2 Phase separation trends

Since the samples being analyzed are water-oil emulsions, water settles down and oil moves up in the liquid column as phase separation occurs. Generally speaking, the middle of the liquid column stayed dispersed for the longest, with the exception of 75W. This corresponds to level L3 sensor, and it was further used to calculate settling trends for 25W, 40W, 50W, and 60W samples for comparison purposes (see **TABLE 3**).

An exponential fit has been used to define the settling curves and trends. For this purpose, MATLAB® Curve Fitting Tool has been used to determine the various coefficients in a second-degree exponential relationship:

$$f = ae^{tb} + ce^{td} \qquad (1)$$

Where f is separated fraction, t is time [s], and a, b, c, d are constants which are different for different emulsions.

**TABLE 3:** SETTLING TREND COEFFICIENTS FOR DIFFERENT SUSPENSIONS

| Emulsion | a | b | c | d |
|---|---|---|---|---|
| 25W | 0.04367 | 8.99E-07 | -0.01505 | -1.48E-05 |
| 40W | 0.05233 | 9.60E-07 | -0.02008 | -1.35E-05 |
| 50W | 0.08373 | 1.01E-06 | -0.04429 | -1.31E-05 |
| 60W | 0.07057 | 1.23E-06 | -0.03241 | -1.56E-05 |

**3.3 Metastable states**

In previous work of Singh et al. [14], it was observed that instead of settling continuously, suspensions went through several metastable states. A similar trend was observed in the present study for all emulsions except 25W, which did not see enough phase separation for the testing period for observable metastable states to occur. The effect was most pronounced at sensor L1.

In a single metastable state, a given emulsion undergoes phase separation at a slower rate, which is followed by phase separation at a significantly faster rate. These metastable states get more pronounced as time progresses (**FIGURE 6**). A possible explanation for this behavior is that the dispersed phase in the freshly prepared emulsion has a very small droplet size but even distribution, with small path length between consecutive droplets. The droplets merge and coalesce easily, but as phases separate out there is more and more path length between consecutive particles. As droplets get larger, they separate out more easily due to gravitational effects overtaking surface tension effects. The slower separation rates correspond to a metastable sub-state where the droplets are coalescing to a larger, "critical" size. As soon as this critical size is reached the phase separation happens rapidly, which corresponds to the next metastable sub-state characterized by faster separation rates. The remaining droplets then have to overcome larger path lengths, which is why metastable states last longer as time progresses.

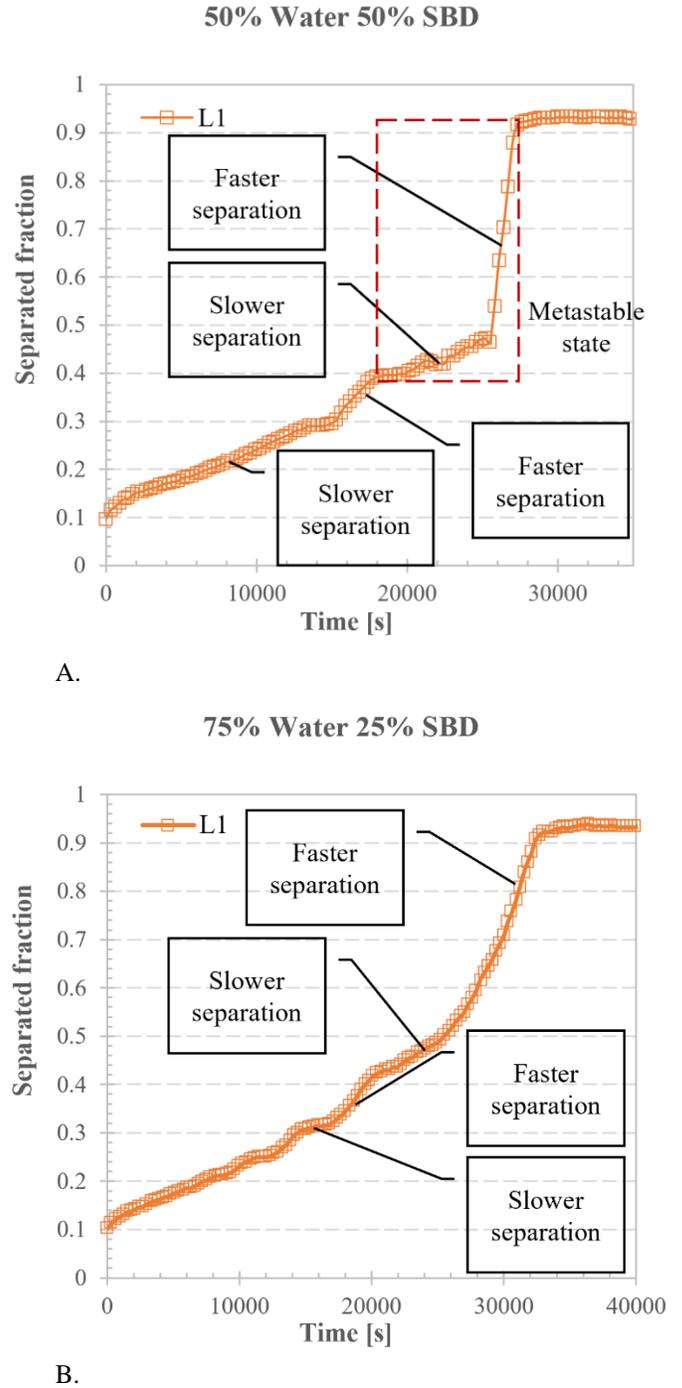

A.

B.

**FIGURE 6:** METASTABLE STATES IN A. 50 W AND B. 75W EMULSIONS AT L1 SENSOR

**4. CONCLUSION**

In present work, fuel-water emulsions from soy biodiesel and distilled water were prepared in different proportions using an ultrasonicator. The emulsions were analyzed for stability and separation characteristics using an experimental setup. The technique presented is non-contact, non-invasive, easy to use and uses low-cost and off-the-shelf components. It was found that as more water was added, the emulsion stability decreased. Phase separation characteristics were found to resemble an exponential relation. Metastable states were seen in the phase separation characteristics, where oil and water phases were seen to separate at consecutive slower and faster separation rates in sub-stages within the same metastable state. It is expected that this method will be explored for stability characterization of other multi-phase liquids as well.

## ACKNOWLEDGEMENTS

This research is funded, in part, by the Mid-America Transportation Center via a grant from the U.S. Department of Transportation's University Transportation Centers Program [USDOT UTC grant number for MATC: 69A3551747107], and this support is gratefully acknowledged. The authors would also like to thank Western Dubuque Biodiesel, LLC for their support. The contents reflect the views of the authors, who are responsible for the facts and the accuracy of the information presented herein and are not necessarily representative of the sponsoring agencies, corporations or persons.

## REFERENCES

[1] S. S. Sazhin et al., "A simple model for puffing/micro-explosions in water-fuel emulsion droplets," *Int. J. Heat Mass Transf.*, vol. 131, no. December, pp. 815–821, 2019.

[2] R. Ocampo-Barrera, R. Villasenor, and A. Diego-Marin, "An experimental study of the effect of water content on combustion of heavy fuel oil/water emulsion droplets," *Combust. Flame*, vol. 126, no. 4, pp. 1845–1855, Sep. 2001.

[3] B. K. Debnath, N. Sahoo, and U. K. Saha, "Adjusting the operating characteristics to improve the performance of an emulsified palm oil methyl ester run diesel engine," *Energy Convers. Manag.*, vol. 69, pp. 191–198, May 2013.

[4] S. S. Reham, H. H. Masjuki, M. A. Kalam, I. Shancita, I. M. Rizwanul Fattah, and A. M. Ruhul, "Study on stability, fuel properties, engine combustion, performance and emission characteristics of biofuel emulsion," *Renew. Sustain. Energy Rev.*, vol. 52, pp. 1566–1579, Dec. 2015.

[5] G. Singh, M. Esmaeilpour, and A. Ratner, "The effect of acetylene black on droplet combustion and flame regime of petrodiesel and soy biodiesel," *Fuel*, vol. 246, 2019.

[6] G. Singh, N. Hentges, D. Johnson, and A. Ratner, "Experimental investigation of combustion behavior of biodiesel-water emulsion (in press)," in *International Mechanical Engineering Congress and Exposition*, 2019.

[7] R. J. Crookes, M. A. A. Nazha, and F. Kiannejad, "Single and Multi Cylinder Diesel-Engine Tests with Vegetable Oil Emulsions," 1992.

[8] L. Yang and K. Du, "A comprehensive review on heat transfer characteristics of TiO2 nanofluids," *International Journal of Heat and Mass Transfer*. 2017.

[9] M. Porras, C. Solans, C. González, and J. M. Gutiérrez, "Properties of water-in-oil (W/O) nano-emulsions prepared by a low-energy emulsification method," *Colloids Surfaces A Physicochem. Eng. Asp.*, vol. 324, no. 1–3, pp. 181–188, Jul. 2008.

[10] C.-Y. Lin and S.-A. Lin, "Effects of emulsification variables on fuel properties of two- and three-phase biodiesel emulsions," *Fuel*, vol. 86, no. 1–2, pp. 210–217, Jan. 2007.

[11] F. Zernike, "Phase contrast, a new method for the microscopic observation of transparent objects," *Physica*, 1942.

[12] C. Anushree and J. Philip, "Assessment of long term stability of aqueous nanofluids using different experimental techniques," *J. Mol. Liq.*, 2016.

[13] G. Singh, S. Pitts, A. Ratner, and E. Lopes, "Settling characteristics of polymeric additives in dodecane," in *ASME International Mechanical Engineering Congress and Exposition, Proceedings (IMECE)*, 2018, vol. 7.

[14] G. Singh, E. Lopes, N. Hentges, D. Becker, and A. Ratner, "Experimental Investigation of the Settling Characteristics of Carbon and Metal Oxide Nanofuels," *J. Nanofluids*, vol. 8, no. 8, 2019.